\documentclass[aps,prl,reprint,groupedaddress]{revtex4-1}
 \usepackage{graphicx}
\begin{document}

%\preprint{}

\title{Experimental Realization of a Nonlinear Acoustic Lens with a Tunable Focus}



\author{Carly M Donahue}
\affiliation{Graduate Aerospace Laboratories (GALCIT), California Institute of Technology, Pasadena, California 91125, USA}
\author{Paul WJ Anzel}
\affiliation{Graduate Aerospace Laboratories (GALCIT), California Institute of Technology, Pasadena, California 91125, USA}
\author{Luca Bonanomi}
\affiliation{Department of Mechanical and Process Engineering (D-MAVT), Swiss Federal Institute of Technology (ETH), Z\"{u}rich, CH}
\author{Thomas A Keller}
\affiliation{Graduate Aerospace Laboratories (GALCIT), California Institute of Technology, Pasadena, California 91125, USA}
\author{Chiara Daraio}
\affiliation{Graduate Aerospace Laboratories (GALCIT), California Institute of Technology, Pasadena, California 91125, USA}
\affiliation{Department of Mechanical and Process Engineering (D-MAVT), Swiss Federal Institute of Technology (ETH), Z\"{u}rich, CH}

%\thanks{}


\date{\today}

\begin{abstract}
We realize a nonlinear acoustic lens composed of a two-dimensional array of sphere chains interfaced with water.  The chains are able to support solitary waves which, when interfaced with a linear medium, transmit compact pulses with minimal oscillations.  When focused, the lens is able to produce compact pressure pulses of high amplitude, the ``sound bullets".  We demonstrate that the focal point can be controlled via pre-compression of the individual chains, as this changes the wave speed within them.  The experimental results agree well both spatially and temporally with analytical predictions over a range of focus locations.    
\end{abstract}

%\keywords{}

\maketitle

%\section{}
%\subsection{}
%\subsubsection{}

\section{Introduction}

\par The focusing of acoustic signals by acoustic lenses is used in a variety of applications such as biomedical imaging or underwater mapping \cite{Kessler1974,korpel1969,Kamgar-Parsi1997}.  Acoustic lenses typically work by using a linear material with a curved geometry that has a speed of sound different from the coupled medium, similar to an optical lens.  Recently, metamaterials and phononic crystals have been used to create flat acoustic lenses based on gradient impedances \cite{Zhang2009} and have achieved an image resolution better than the diffraction limit \cite{Sukhovich2009} or subwavelength resolution \cite{Zhu2010,Christensen2012}.  An acoustic lens composed of a nonlinear material offers a novel method of filtering and focusing signals \cite{Spadoni2010}.  Granular crystals, defined as chains or lattices of particles in contact with each other, are a widely studied example of nonlinear media   \cite{Nesterenko2001,Sen2008}.  Their nonlinearity arises from the Hertzian contact between particles; for spheres, the restitution force is proportional to the relative displacement to the three-halves power.  When struck, granular crystals can support solitary waves, which are stable pulses with a compact profile and a speed nonlinearly dependent on their amplitude.  Unlike linear materials, no speed of sound is defined for a strongly nonlinear granular crystal; instead speed of the solitary wave is instead determined by the material properties of the individual granules, the striking force, and the amount that the particles are pre-compressed prior to actuation \cite{Coste1999,Daraio2006}.  Granular crystals were first investigated for use in an acoustic lens by Spadoni and Daraio \cite{Spadoni2010}.  In their study, a one-dimensional array of granular chains was used to focus stresses in a polymeric plate and the resulting stress was imaged using high-speed photoelasticity. 

\par  In this work, we have developed a nonlinear lens composed of a two-dimensional array of sphere chains interfaced with water, and we image the focused pressure pulses using a scanning hydrophone.  We show the ability to focus acoustic signals in different locations in water by tuning the static pre-compression applied to the individual granular chains.  The solitary waves formed within a two-dimensional array of chains are transmitted in the water as a ``sound bullet", a bipolar focused region.  A sound bullet is illustrated in Figure \ref{Experimental Setup}a where the surfaces are contours of constant pressure.  Here, the positive and negative regions of the sound bullet are shown to be compact in three-dimensions.  
\par The nonlinear acoustic lens has potential advantages over linear lenses.  First, each element of the nonlinear lens generates pulses with minimal oscillations, so the focus is compact, and other areas of constructive interference are minimized.  Second, the lens can support and focus high-amplitude solitary waves generated by high strength actuators.  Finally, because the solitary wave speed depends on pre-compression \cite{Daraio2006}, the focal length can be controlled through mechanical means.
\par The unique acoustic properties of the sound bullets have many possible applications to ultrasonic and underwater acoustics in which transducers for the generation of compact, high-amplitude pulses continue to be studied \cite{Baac2012}.  High Intensity Focused Ultrasound (HIFU) for therapeutic applications requires large pressure amplitudes at the focus, while minimizing the amplitudes elsewhere to prevent damage to healthy tissue \cite{Pajek2012, Kennedy2003}.  For generating high resolution acoustic images, short cycles are desired \cite{Szabo2013}.  Since the timing and therefore the position of the focus can be controlled mechanically, the lens could be designed without the use of complicated electronics thus operate in particularly inhospitable conditions \cite{Brie1998}.  
	
\section{Experimental Setup}

 \begin{figure*}[h!]
   \centering
     \includegraphics[width = \textwidth]{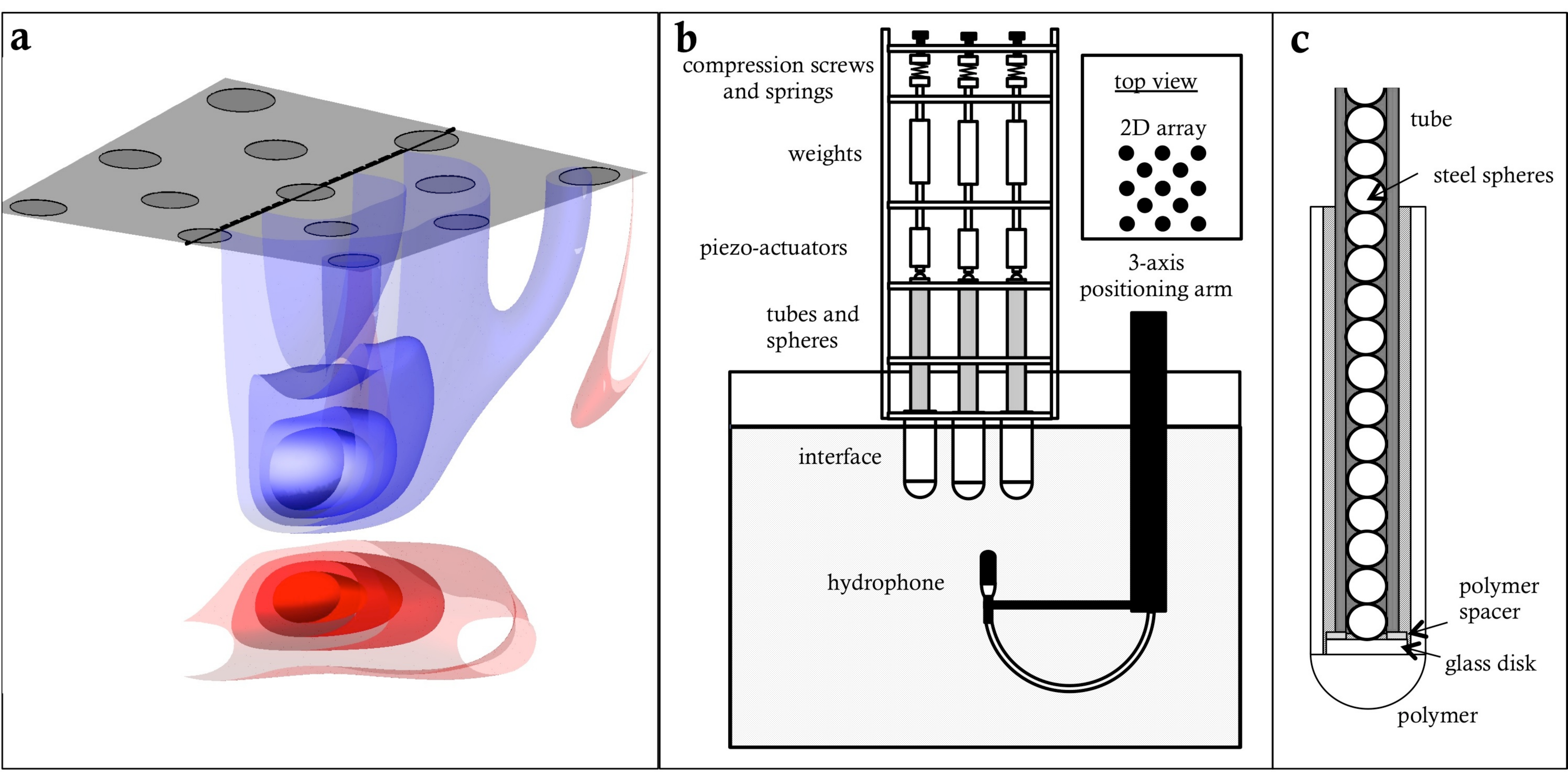}
 \caption{(a) Surfaces of constant pressure for a sound bullet generated using a 2D nonlinear acoustic lens.  The red surfaces are at 20, 40, 60, and 80 percent of the peak positive pressure and the blue surfaces are respectively for negative pressure.   (b) Schematic of the experimental setup. (c) A cross section of the interface supporting and separating granular chain from the water.  \label{Experimental Setup}}

 \end{figure*} 
	
\par The experimental setup is shown in Figure \ref{Experimental Setup}b.  The lens is composed of 13 independent granular chains arranged in a square lattice (Figure \ref{Experimental Setup}b inset) such that the nearest neighbor chains are 5.6 cm apart.  Each chain has a total of 30 stainless steel spheres of grade 440C and 9.296 mm in diameter.  
%The spheres are held by a steel tube with an inner diameter of 9.347 mm $\pm$ 0.025 mm.  

\par One crucial component of the nonlinear acoustic lens is the interface which holds the spheres and separates them from the water.  Studies of a granular chain interacting with a boundary served to guide the design \cite{Job2005,Falcon1998,Yang2011,Yang2012}, but most of the prior studies have concentrated on the reflected solitary waves and not the wave transmitted though the boundary.   We found that the properties of the focus are sensitive to the design of the interface.  
\par The interface designed for the chains has three components (Figure \ref{Experimental Setup}c): a glass disk, a polymer encasing, and a spacer.  The collision time of the last sphere with the interface depends on the material properties of the interface, such that softer materials result in longer collision times and, therefore, larger pulse widths \cite{Yang2011}.  Consequently, the last sphere is in contact with a stiff borosilicate glass disk, diameter of 17.6 mm and thickness of 3.2 mm.  The borosilicate glass has a density of 2210 kg/m$^3$ and an elastic modulus of 63 GPa \cite{MC}.  The chains are individually supported by a polymer casing created using an Objet 3D printer.  The side tube of the polymer encasing is made with VeroClear, a stiff polymer, and has a radius of 11.7 mm, thickness of 2.7 mm and length of 97.8 mm.  VeroClear has a density of 1160 kg/m$^3$, an elastic modulus of 1.13 GPa, and a poisson's ratio of 0.38 as measured using an Olympus V112 longitudinal transducer and an Olympus V156 shear transducer.  The bottom hemisphere of radius 11.7 mm is made with FLX9985-DM, a rubber-like polymer with a Shore A hardness between 80-85 MPa \cite{Objet}.  The polymer encasing serves three purposes, (a) to support the glass disk, (b) to waterproof the chains, and (c) to damp resonant vibrations of the glass disk.  The interface also contains a small spacer made of FLX9985-DM between the steel tube holding the spheres and the glass disk, which serves (a) to ensure an adequate spacing between the two and (b) further reduce any resonant vibrations of the glass disk. Oscillations of the polymer casing are additionally damped by securing a thin neoprene rubber sheet with zip ties to the outside. 

\par The chains are pulsed with Piezomechanik piezo-actuators (PIA 1000/10/15 PL16).  Each piezo-actuator is connected to a separate high voltage pulser (IXYS HV1000) such that the pulsers keep the piezos at a high voltage of 715 V for 20 $\mu$s.   The solitary wave formed within the chain is probed by a Polytec OFV-505 laser vibrometer.  The maximum velocity of a sphere five particles from the bottom is 0.555 m/s $\pm$ 0.033 m/s, which leads to a solitary wave speed of 643 m/s and a maximum force between the particles of 158 N \cite{Nesterenko2001}. 

\par Measurements are made in a tank of dimensions 1 m long by 0.75 m wide by 0.75 m tall manufactured by Precision Acoustics.  The tank is filled with deionized water at a temperature of 22 $^\circ$C.  Attached to the tank is a three-axis positioning arm on which the hydrophone is mounted.  The hydrophone is manufactured by Bruel and Kjaer type 8103 and it has a frequency range of 0.1 Hz to 180 kHz and a diameter of 9.5 mm.  

\par The time at which the signal from each chain reaches the water is controlled by pre-compression of the chain by a set of screws and springs.  The screws are able to produce a force on the spheres between 0 - 70 N, which is able to generate a change in the arrival time of the signal to the water by approximately 0 - 55 $\mu$s.   The maximum amount of force is limited by the structural strength of the polymer interface.  The focal location is tuned empirically; specifically, the hydrophone is placed at the desired focal location, and each piezo is pulsed individually.  The screws are then tightened until the desired time-of-flight is achieved (the time between triggering a piezo to when the peak pressure is measured at the hydrophone) such that the time-of-flight for all elements are equal. 

\section{Theoretical Description}

\par The theoretical model used to describe the focusing of the nonlinear acoustic lens was developed in \cite{Spadoni2010} and is briefly described here.  In the long wavelength limit, the dynamics of a wave traveling strongly nonlinear granular chain can be described by \cite{Nesterenko2001} 
 \begin{equation}
 \label{solitarywave}
   v_n(t) = \left\{
     \begin{array}{lr}
       A_n\cos^4(\alpha_n) & \alpha_n \in [-\pi/2,\pi/2]\\
       0 & otherwise
     \end{array}
   \right.
\end{equation} 
where
\begin{displaymath}
\alpha_n = k_s[x-V_{s,n}(t-\Delta t_n)].
\end{displaymath}
Here, $v_n$ is the velocity corresponding to the $n^{th}$ chain, $A_n$ is the amplitude, $x$ is the position, $t$ is the time, $\Delta t_n$ is the phase delay, $k_s = \sqrt{10}/(5D)$ is the wavenumber with $D$ equal to the diameter of the spheres, and $V_{s,n}$ is the solitary wave velocity.  The solitary wave speed is related to the static pre-compression, $F_0$, and the maximum force, $F_m$, by
 \begin{equation}
 \label{solitarywavespped}
V_{s,n} \propto \frac{F_0^{1/6}}{f_r^{2/3}-1}\left(3+2f_r^{5/3}-5f_r^{2/3}\right)^{1/2},
 \end{equation}
 where $f_r = F_0/F_m$.

\par To derive the pressure in the fluid target, the motion of the interface is assumed to have the same waveform as the solitary wave velocity such that \cite{Spadoni2010}
\begin{equation}
\label{pressure}
p(x,y,t) = \frac{\rho c k_s}{2} \sum^N_{n=1} \frac{B_n V_{s,n} [2 \sin (2 k_s \phi_n) + \sin(4 k_s \phi_n)]}{r_n}.
\end{equation}
Here, the pressure field is nonzero for $k_s\phi_n \in [-\pi/2, \pi/2]$, $\rho$ is the density of the target, $c$ is the speed of sound in the target, $B_n$ is chosen to match the pressure amplitude, $r_n$ is the distance from the chain, and $\phi_n = V_{s,n}(r_n/c-t+\Delta t_n)$.

\section{Results}

  %---FIGURE
 \begin{figure*}[h!]
   \centering
     \includegraphics[width = \textwidth]{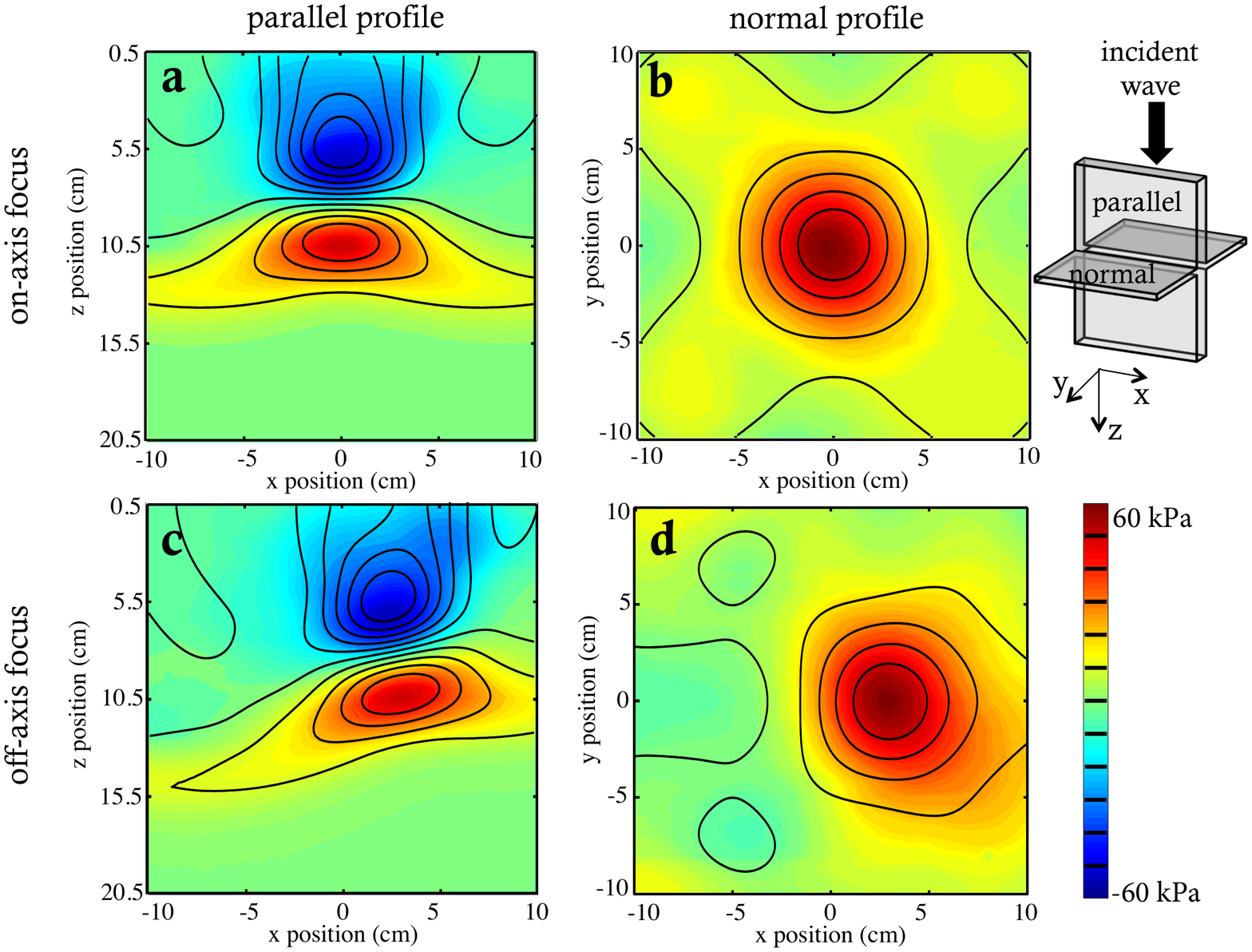}
 \caption{Pressure map showing the formation of a sound bullet for (a, b) a position along the axis of the lens 10 cm away  and (c, d) a position 3 cm off-axis and 10 cm away from the lens.   (a, c) show the cross sections parallel to the incident wave, and (b, d) show the cross sections normal to the incident wave.  The black lines are contour lines of the theoretical prediction.  \label{pressuremap}}  
 \end{figure*}
 \begin{figure*}[h!]
   \centering
     \includegraphics[width = \columnwidth]{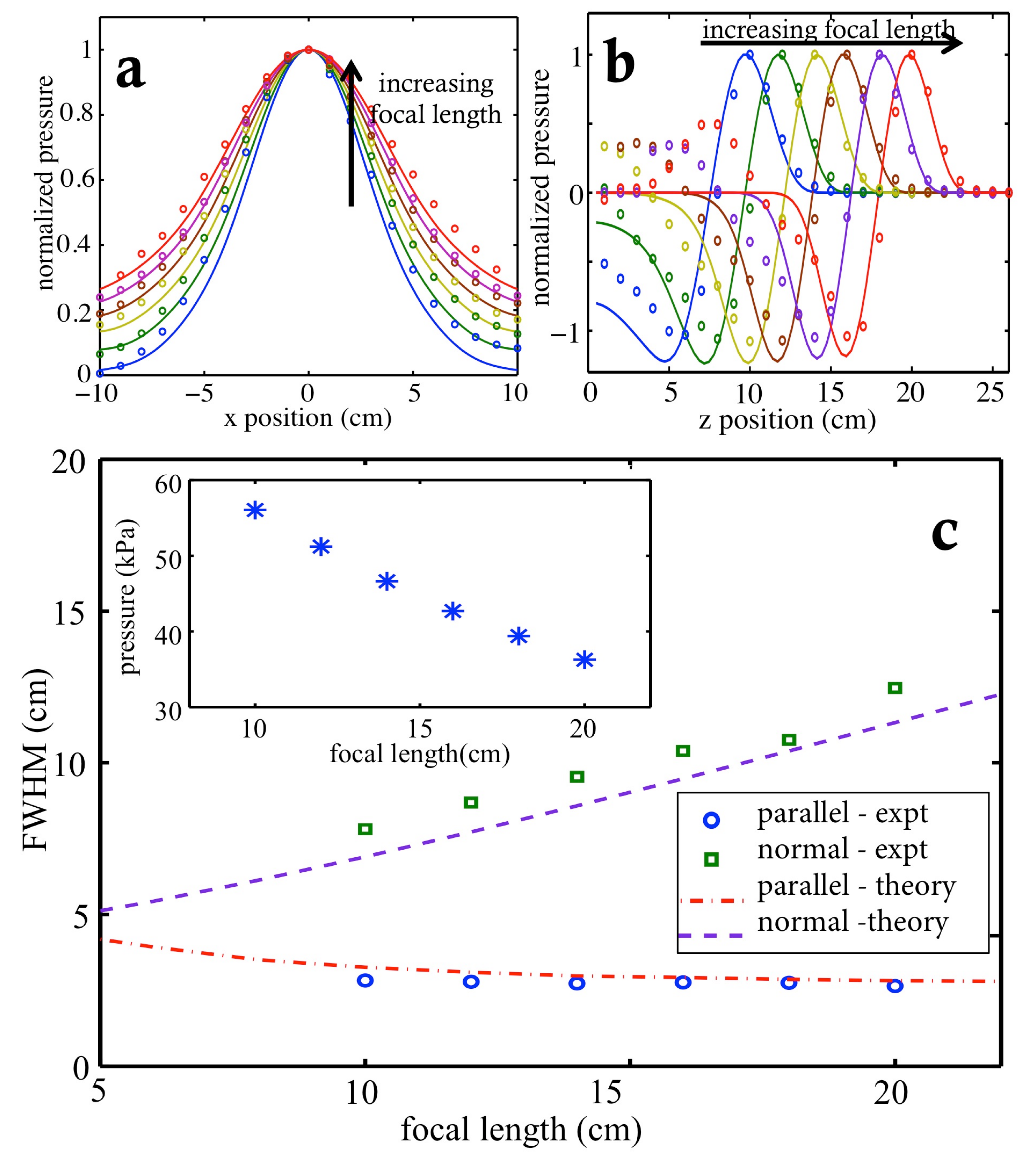}
 \caption{Normalized pressure on a line that intersects the focus (a) normal to the incident wave (along the x-axis) and (b) parallel to the incident wave (along the z-axis) for experiment (circles) and theory (lines) for focal lengths between 10 cm and 20 cm.  (c) FWHM calculated from (a,b).  The inset of (c) shows peak pressure at the focus. \label{comparison}
}
 \end{figure*} 

 \begin{figure*}[h!]
   \centering
     \includegraphics[width = \columnwidth]{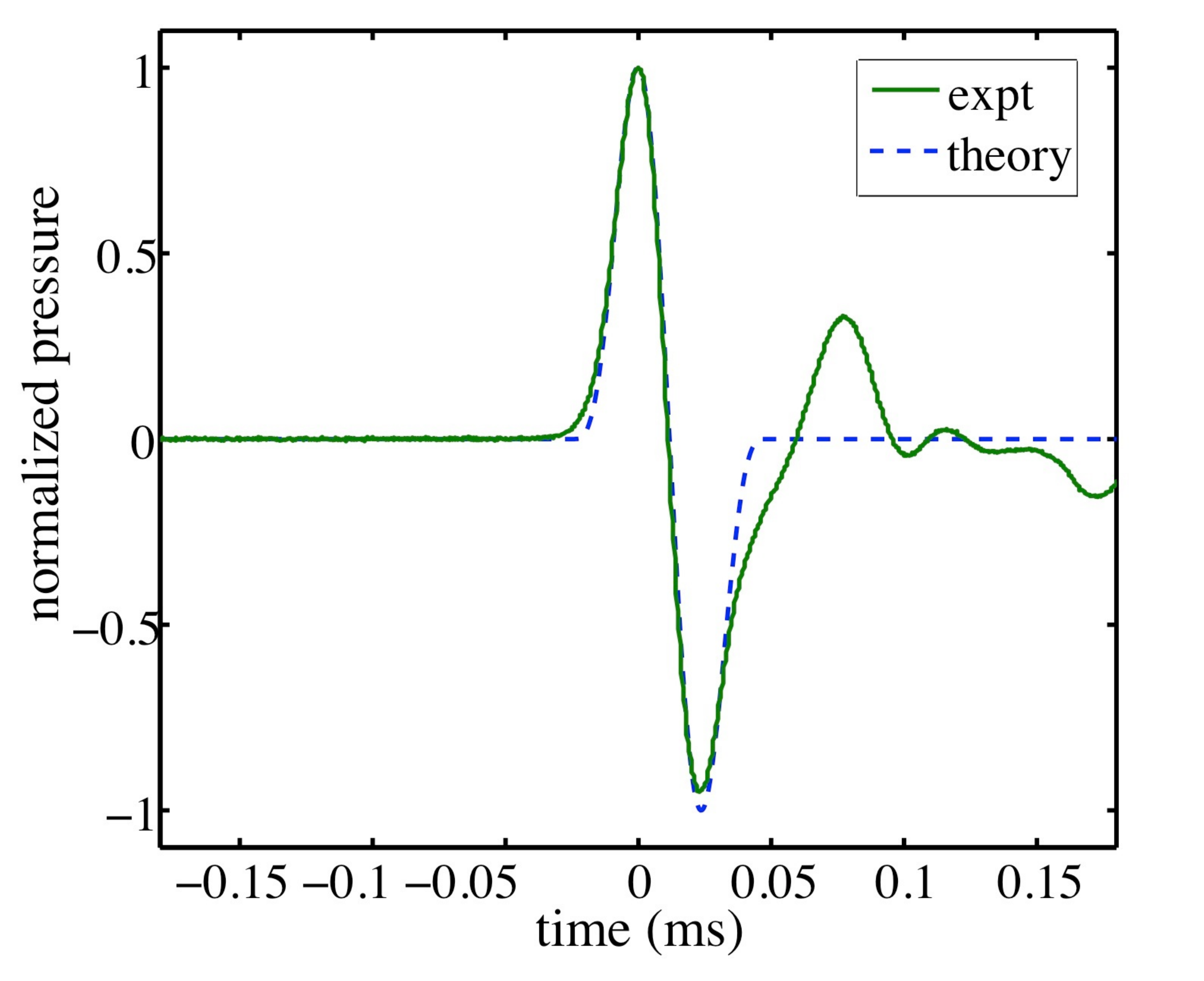}
 \caption{ Pressure profile at the focus for an on-axis focus at a focal length of 10 cm. \label{time}
}
 \end{figure*}

\par Measurements of the sound bullet are made for a variety of focal locations by controlling the time delay of the signals generated by each chain.  Figure \ref{pressuremap} shows the pressure maps for both an on-axis (the sound bullet located along the symmetrical axis of the lens) and off-axis as the peak pressure of the sound bullet arrives at the focus.  The parallel profile shows the large positive pressure pole at the focus with the negative pole trailing behind; whereas the normal profiles shows a cross section of just the positive pole.  Notice that outside of the focal region, the pressure is small.  The theoretical predictions (black lines) have an excellent agreement with experiment.  The asymmetrical deviations of the sound bullet from the theory in Figures \ref{pressuremap}b and c is due to differences in signal strength from individual chains as a result of differences of the firing strengths of the piezo-actuators and in the manufacture of the interface.

%($x = 0$ cm, $y = 0$ cm, and $z = 10$ cm)
%($x = 3$ cm, $y = 0$ cm, and $z = 10$ cm).
 
\par To further investigate the ability of the theory to capture the spatial form of the experiment, the results are analyzed for different on-axis focal lengths.  Figures \ref{comparison}a and b show the pressure along a line that intersects the focus normal and parallel, respectively, to the incident wave.  Here, the pressure is normalized to the maximum pressure at the focus.  From the data presented in Figures \ref{comparison}a and b, the experimental data is fitted to find the full width half maximum (FWHM) of the positive pressure pole for both parallel and normal profiles as shown in Figure \ref{comparison}c.  Excellent agreement is achieved between experiment and theory as the on-axis focal length is increased, showing that the theory is robust for a range of focal locations.  The amplitude of the sound bullet is analyzed as a function of the focal length in the inset of Figure \ref{comparison}c.  The amplitude decreases by approximately one-third as the focal length is doubled.  

\par The shape of the sound bullet depends on the position of the sound bullet and the placement of the chains.  As seen in Figure \ref{comparison}a, as the focal length decreases, the aspect ratio in the normal and parallel direction approaches one, resulting in a``rounder" sound bullet.   Additionally, the four-fold symmetry of Figure \ref{pressuremap}b is due to the square packing of the chains; a hexagonal packing would result in a focus with six-fold symmetry.   The size of the sound bullet is dictated by the wavelength of the solitary waves within each chain.  The wavelength of the solitary waves is approximately equal to $5D$, therefore, reducing the size of the spheres will result in a smaller focus.

\par The temporal profiles of the focused signals are shown in Figure \ref{time}.  The theory is able to capture both the form and the wavelength of the bipolar pulse.  However, the experimental profile shows some oscillations after the negative pole.  Such oscillations are a result of resonant vibrations of the interface.  Further investigation and development of the interface can lead to a further decrease in the additional oscillations.

 %---FIGURE

\section{Conclusion}

We have designed and constructed a two-dimensional nonlinear acoustic lens interfaced with water.  We have demonstrated the tunability of the focal position by varying the static pre-compression applied to the individual chains.  The experimental data agrees well with the theory.  The nonlinear acoustic lens could find applications in underwater and ultrasonic imaging, hydrophone calibration, and high-intensity focusing.  

\par This work is supported by the Office of Naval Research, YIP program, Award Number N000141010718.  Paul Anzel was supported by a NASA Office of the Chief TechnologistÕs Space Technology Research Fellowship Number NNX11AN65H.  We also thank Thevamaran Ramathasan for material property measurements and Wei-Hsun Lin for programming support.  

\bibliography{MyCollection.bib}

\begin{thebibliography}{23}%
\makeatletter
\providecommand \@ifxundefined [1]{%
 \@ifx{#1\undefined}
}%
\providecommand \@ifnum [1]{%
 \ifnum #1\expandafter \@firstoftwo
 \else \expandafter \@secondoftwo
 \fi
}%
\providecommand \@ifx [1]{%
 \ifx #1\expandafter \@firstoftwo
 \else \expandafter \@secondoftwo
 \fi
}%
\providecommand \natexlab [1]{#1}%
\providecommand \enquote  [1]{``#1''}%
\providecommand \bibnamefont  [1]{#1}%
\providecommand \bibfnamefont [1]{#1}%
\providecommand \citenamefont [1]{#1}%
\providecommand \href@noop [0]{\@secondoftwo}%
\providecommand \href [0]{\begingroup \@sanitize@url \@href}%
\providecommand \@href[1]{\@@startlink{#1}\@@href}%
\providecommand \@@href[1]{\endgroup#1\@@endlink}%
\providecommand \@sanitize@url [0]{\catcode `\\12\catcode `\$12\catcode
  `\&12\catcode `\#12\catcode `\^12\catcode `\_12\catcode `\%12\relax}%
\providecommand \@@startlink[1]{}%
\providecommand \@@endlink[0]{}%
\providecommand \url  [0]{\begingroup\@sanitize@url \@url }%
\providecommand \@url [1]{\endgroup\@href {#1}{\urlprefix }}%
\providecommand \urlprefix  [0]{URL }%
\providecommand \Eprint [0]{\href }%
\providecommand \doibase [0]{http://dx.doi.org/}%
\providecommand \selectlanguage [0]{\@gobble}%
\providecommand \bibinfo  [0]{\@secondoftwo}%
\providecommand \bibfield  [0]{\@secondoftwo}%
\providecommand \translation [1]{[#1]}%
\providecommand \BibitemOpen [0]{}%
\providecommand \bibitemStop [0]{}%
\providecommand \bibitemNoStop [0]{.\EOS\space}%
\providecommand \EOS [0]{\spacefactor3000\relax}%
\providecommand \BibitemShut  [1]{\csname bibitem#1\endcsname}%
\let\auto@bib@innerbib\@empty
%</preamble>
\bibitem [{\citenamefont {Kessler}(1974)}]{Kessler1974}%
  \BibitemOpen
  \bibfield  {author} {\bibinfo {author} {\bibfnamefont {L.~W.}\ \bibnamefont
  {Kessler}},\ }\href@noop {} {\bibfield  {journal} {\bibinfo  {journal}
  {Journal of the Acoustic Society of America}\ }\textbf {\bibinfo {volume}
  {55}},\ \bibinfo {pages} {909} (\bibinfo {year} {1974})}\BibitemShut
  {NoStop}%
\bibitem [{\citenamefont {Korpel}(1969)}]{korpel1969}%
  \BibitemOpen
  \bibfield  {author} {\bibinfo {author} {\bibfnamefont {A.}~\bibnamefont
  {Korpel}},\ }\href {\doibase 10.1109/JQE.1969.1081932} {\bibfield  {journal}
  {\bibinfo  {journal} {IEEE Journal of Quantum Electronics}\ }\textbf
  {\bibinfo {volume} {5}},\ \bibinfo {pages} {324} (\bibinfo {year}
  {1969})}\BibitemShut {NoStop}%
\bibitem [{\citenamefont {Kamgar-Parsi}\ \emph {et~al.}(1997)\citenamefont
  {Kamgar-Parsi}, \citenamefont {Johnson}, \citenamefont {Folds},\ and\
  \citenamefont {Belcher}}]{Kamgar-Parsi1997}%
  \BibitemOpen
  \bibfield  {author} {\bibinfo {author} {\bibfnamefont {B.}~\bibnamefont
  {Kamgar-Parsi}}, \bibinfo {author} {\bibfnamefont {B.}~\bibnamefont
  {Johnson}}, \bibinfo {author} {\bibfnamefont {D.~L.}\ \bibnamefont {Folds}},
  \ and\ \bibinfo {author} {\bibfnamefont {E.~O.}\ \bibnamefont {Belcher}},\
  }\href {\doibase 10.1002/(SICI)1098-1098(1997)8:4<377::AID-IMA4>3.0.CO;2-7}
  {\bibfield  {journal} {\bibinfo  {journal} {International Journal of Imaging
  Systems and Technology}\ }\textbf {\bibinfo {volume} {8}},\ \bibinfo {pages}
  {377} (\bibinfo {year} {1997})}\BibitemShut {NoStop}%
\bibitem [{\citenamefont {Zhang}\ \emph {et~al.}(2009)\citenamefont {Zhang},
  \citenamefont {Yin},\ and\ \citenamefont {Fang}}]{Zhang2009}%
  \BibitemOpen
  \bibfield  {author} {\bibinfo {author} {\bibfnamefont {S.}~\bibnamefont
  {Zhang}}, \bibinfo {author} {\bibfnamefont {L.}~\bibnamefont {Yin}}, \ and\
  \bibinfo {author} {\bibfnamefont {N.}~\bibnamefont {Fang}},\ }\href {\doibase
  10.1103/PhysRevLett.102.194301} {\bibfield  {journal} {\bibinfo  {journal}
  {Physical Review Letters}\ }\textbf {\bibinfo {volume} {102}},\ \bibinfo
  {pages} {1} (\bibinfo {year} {2009})}\BibitemShut {NoStop}%
\bibitem [{\citenamefont {Sukhovich}\ \emph {et~al.}(2009)\citenamefont
  {Sukhovich}, \citenamefont {Merheb}, \citenamefont {Muralidharan},
  \citenamefont {Vasseur}, \citenamefont {Pennec}, \citenamefont {Deymier},\
  and\ \citenamefont {Page}}]{Sukhovich2009}%
  \BibitemOpen
  \bibfield  {author} {\bibinfo {author} {\bibfnamefont {A.}~\bibnamefont
  {Sukhovich}}, \bibinfo {author} {\bibfnamefont {B.}~\bibnamefont {Merheb}},
  \bibinfo {author} {\bibfnamefont {K.}~\bibnamefont {Muralidharan}}, \bibinfo
  {author} {\bibfnamefont {J.}~\bibnamefont {Vasseur}}, \bibinfo {author}
  {\bibfnamefont {Y.}~\bibnamefont {Pennec}}, \bibinfo {author} {\bibfnamefont
  {P.}~\bibnamefont {Deymier}}, \ and\ \bibinfo {author} {\bibfnamefont
  {J.}~\bibnamefont {Page}},\ }\href {\doibase 10.1103/PhysRevLett.102.154301}
  {\bibfield  {journal} {\bibinfo  {journal} {Physical Review Letters}\
  }\textbf {\bibinfo {volume} {102}},\ \bibinfo {pages} {1} (\bibinfo {year}
  {2009})}\BibitemShut {NoStop}%
\bibitem [{\citenamefont {Zhu}\ \emph {et~al.}(2010)\citenamefont {Zhu},
  \citenamefont {Christensen}, \citenamefont {Jung}, \citenamefont {Yin},
  \citenamefont {Fok},\ and\ \citenamefont {Zhang}}]{Zhu2010}%
  \BibitemOpen
  \bibfield  {author} {\bibinfo {author} {\bibfnamefont {J.}~\bibnamefont
  {Zhu}}, \bibinfo {author} {\bibfnamefont {J.}~\bibnamefont {Christensen}},
  \bibinfo {author} {\bibfnamefont {J.}~\bibnamefont {Jung}}, \bibinfo {author}
  {\bibfnamefont {X.}~\bibnamefont {Yin}}, \bibinfo {author} {\bibfnamefont
  {L.}~\bibnamefont {Fok}}, \ and\ \bibinfo {author} {\bibfnamefont
  {X.}~\bibnamefont {Zhang}},\ }\href {\doibase 10.1038/nphys1804} {\bibfield
  {journal} {\bibinfo  {journal} {Nature Physics}\ }\textbf {\bibinfo {volume}
  {7}},\ \bibinfo {pages} {52} (\bibinfo {year} {2010})}\BibitemShut {NoStop}%
\bibitem [{\citenamefont {Christensen}\ and\ \citenamefont
  {de~Abajo}(2012)}]{Christensen2012}%
  \BibitemOpen
  \bibfield  {author} {\bibinfo {author} {\bibfnamefont {J.}~\bibnamefont
  {Christensen}}\ and\ \bibinfo {author} {\bibfnamefont {F.~J.~G.}\
  \bibnamefont {de~Abajo}},\ }\href {\doibase 10.1103/PhysRevLett.108.124301}
  {\bibfield  {journal} {\bibinfo  {journal} {Physical Review Letters}\
  }\textbf {\bibinfo {volume} {108}},\ \bibinfo {pages} {124301} (\bibinfo
  {year} {2012})}\BibitemShut {NoStop}%
\bibitem [{\citenamefont {Spadoni}\ and\ \citenamefont
  {Daraio}(2010)}]{Spadoni2010}%
  \BibitemOpen
  \bibfield  {author} {\bibinfo {author} {\bibfnamefont {A.}~\bibnamefont
  {Spadoni}}\ and\ \bibinfo {author} {\bibfnamefont {C.}~\bibnamefont
  {Daraio}},\ }\href {http://www.ncbi.nlm.nih.gov/pubmed/20368461} {\bibfield
  {journal} {\bibinfo  {journal} {Proceedings of the National Academy of
  Sciences of the United States of America}\ }\textbf {\bibinfo {volume}
  {107}},\ \bibinfo {pages} {7230} (\bibinfo {year} {2010})}\BibitemShut
  {NoStop}%
\bibitem [{\citenamefont {Nesterenko}(2001)}]{Nesterenko2001}%
  \BibitemOpen
  \bibfield  {author} {\bibinfo {author} {\bibfnamefont {V.~F.}\ \bibnamefont
  {Nesterenko}},\ }\href@noop {} {\emph {\bibinfo {title} {{Dynamics of
  Heterogenous Materials}}}}\ (\bibinfo  {publisher} {Springer},\ \bibinfo
  {year} {2001})\BibitemShut {NoStop}%
\bibitem [{\citenamefont {Sen}\ \emph {et~al.}(2008)\citenamefont {Sen},
  \citenamefont {Hong}, \citenamefont {Bang}, \citenamefont {Avalos},\ and\
  \citenamefont {Doney}}]{Sen2008}%
  \BibitemOpen
  \bibfield  {author} {\bibinfo {author} {\bibfnamefont {S.}~\bibnamefont
  {Sen}}, \bibinfo {author} {\bibfnamefont {J.}~\bibnamefont {Hong}}, \bibinfo
  {author} {\bibfnamefont {J.}~\bibnamefont {Bang}}, \bibinfo {author}
  {\bibfnamefont {E.}~\bibnamefont {Avalos}}, \ and\ \bibinfo {author}
  {\bibfnamefont {R.}~\bibnamefont {Doney}},\ }\href {\doibase
  10.1016/j.physrep.2007.10.007} {\bibfield  {journal} {\bibinfo  {journal}
  {Physics Reports}\ }\textbf {\bibinfo {volume} {462}},\ \bibinfo {pages} {21}
  (\bibinfo {year} {2008})}\BibitemShut {NoStop}%
\bibitem [{\citenamefont {Coste}\ and\ \citenamefont
  {Gilles}(1999)}]{Coste1999}%
  \BibitemOpen
  \bibfield  {author} {\bibinfo {author} {\bibfnamefont {C.}~\bibnamefont
  {Coste}}\ and\ \bibinfo {author} {\bibfnamefont {B.}~\bibnamefont {Gilles}},\
  }\href {\doibase 10.1007/s100510050598} {\bibfield  {journal} {\bibinfo
  {journal} {European Physical Journal B}\ }\textbf {\bibinfo {volume} {7}},\
  \bibinfo {pages} {155} (\bibinfo {year} {1999})}\BibitemShut {NoStop}%
\bibitem [{\citenamefont {Daraio}\ \emph {et~al.}(2006)\citenamefont {Daraio},
  \citenamefont {Nesterenko}, \citenamefont {Herbold},\ and\ \citenamefont
  {Jin}}]{Daraio2006}%
  \BibitemOpen
  \bibfield  {author} {\bibinfo {author} {\bibfnamefont {C.}~\bibnamefont
  {Daraio}}, \bibinfo {author} {\bibfnamefont {V.}~\bibnamefont {Nesterenko}},
  \bibinfo {author} {\bibfnamefont {E.}~\bibnamefont {Herbold}}, \ and\
  \bibinfo {author} {\bibfnamefont {S.}~\bibnamefont {Jin}},\ }\href {\doibase
  10.1103/PhysRevE.73.026610} {\bibfield  {journal} {\bibinfo  {journal}
  {Physical Review E}\ }\textbf {\bibinfo {volume} {73}},\ \bibinfo {pages} {1}
  (\bibinfo {year} {2006})}\BibitemShut {NoStop}%
\bibitem [{\citenamefont {Baac}\ \emph {et~al.}(2012)\citenamefont {Baac},
  \citenamefont {Ok}, \citenamefont {Maxwell}, \citenamefont {Lee},
  \citenamefont {Chen}, \citenamefont {Hart}, \citenamefont {Xu}, \citenamefont
  {Yoon},\ and\ \citenamefont {Guo}}]{Baac2012}%
  \BibitemOpen
  \bibfield  {author} {\bibinfo {author} {\bibfnamefont {H.~W.}\ \bibnamefont
  {Baac}}, \bibinfo {author} {\bibfnamefont {J.~G.}\ \bibnamefont {Ok}},
  \bibinfo {author} {\bibfnamefont {A.}~\bibnamefont {Maxwell}}, \bibinfo
  {author} {\bibfnamefont {K.-T.}\ \bibnamefont {Lee}}, \bibinfo {author}
  {\bibfnamefont {Y.-C.}\ \bibnamefont {Chen}}, \bibinfo {author}
  {\bibfnamefont {A.~J.}\ \bibnamefont {Hart}}, \bibinfo {author}
  {\bibfnamefont {Z.}~\bibnamefont {Xu}}, \bibinfo {author} {\bibfnamefont
  {E.}~\bibnamefont {Yoon}}, \ and\ \bibinfo {author} {\bibfnamefont {L.~J.}\
  \bibnamefont {Guo}},\ }\href {\doibase 10.1038/srep00989} {\bibfield
  {journal} {\bibinfo  {journal} {Scientific reports}\ }\textbf {\bibinfo
  {volume} {2}},\ \bibinfo {pages} {989} (\bibinfo {year} {2012})}\BibitemShut
  {NoStop}%
\bibitem [{\citenamefont {Pajek}\ and\ \citenamefont
  {Hynynen}(2012)}]{Pajek2012}%
  \BibitemOpen
  \bibfield  {author} {\bibinfo {author} {\bibfnamefont {D.}~\bibnamefont
  {Pajek}}\ and\ \bibinfo {author} {\bibfnamefont {K.}~\bibnamefont
  {Hynynen}},\ }\href@noop {} {\bibfield  {journal} {\bibinfo  {journal}
  {Acoustics Today}\ }\textbf {\bibinfo {volume} {8}} (\bibinfo {year}
  {2012})}\BibitemShut {NoStop}%
\bibitem [{\citenamefont {Kennedy}(2003)}]{Kennedy2003}%
  \BibitemOpen
  \bibfield  {author} {\bibinfo {author} {\bibfnamefont {J.~E.}\ \bibnamefont
  {Kennedy}},\ }\href {\doibase 10.1259/bjr/17150274} {\bibfield  {journal}
  {\bibinfo  {journal} {British Journal of Radiology}\ }\textbf {\bibinfo
  {volume} {76}},\ \bibinfo {pages} {590} (\bibinfo {year} {2003})}\BibitemShut
  {NoStop}%
\bibitem [{\citenamefont {Szabo}(2004)}]{Szabo2013}%
  \BibitemOpen
  \bibfield  {author} {\bibinfo {author} {\bibfnamefont {T.~L.}\ \bibnamefont
  {Szabo}},\ }\href@noop {} {\emph {\bibinfo {title} {{Diagnostic Ultrasound
  Imaging: Inside Out}}}}\ (\bibinfo  {publisher} {Academic Press},\ \bibinfo
  {year} {2004})\BibitemShut {NoStop}%
\bibitem [{\citenamefont {Brie}\ \emph {et~al.}(1998)\citenamefont {Brie},
  \citenamefont {Hoyle}, \citenamefont {Codazzi}, \citenamefont {Hsu},\ and\
  \citenamefont {Mueller}}]{Brie1998}%
  \BibitemOpen
  \bibfield  {author} {\bibinfo {author} {\bibfnamefont {A.}~\bibnamefont
  {Brie}}, \bibinfo {author} {\bibfnamefont {D.}~\bibnamefont {Hoyle}},
  \bibinfo {author} {\bibfnamefont {D.}~\bibnamefont {Codazzi}}, \bibinfo
  {author} {\bibfnamefont {K.}~\bibnamefont {Hsu}}, \ and\ \bibinfo {author}
  {\bibfnamefont {M.~C.}\ \bibnamefont {Mueller}},\ }\href@noop {} {\bibfield
  {journal} {\bibinfo  {journal} {Oilfield Review}\ } (\bibinfo {year}
  {1998})}\BibitemShut {NoStop}%
\bibitem [{\citenamefont {Job}\ \emph {et~al.}(2005)\citenamefont {Job},
  \citenamefont {Melo}, \citenamefont {Sokolow},\ and\ \citenamefont
  {Sen}}]{Job2005}%
  \BibitemOpen
  \bibfield  {author} {\bibinfo {author} {\bibfnamefont {S.}~\bibnamefont
  {Job}}, \bibinfo {author} {\bibfnamefont {F.}~\bibnamefont {Melo}}, \bibinfo
  {author} {\bibfnamefont {A.}~\bibnamefont {Sokolow}}, \ and\ \bibinfo
  {author} {\bibfnamefont {S.}~\bibnamefont {Sen}},\ }\href {\doibase
  10.1103/PhysRevLett.94.178002} {\bibfield  {journal} {\bibinfo  {journal}
  {Physical Review Letters}\ }\textbf {\bibinfo {volume} {94}},\ \bibinfo
  {pages} {1} (\bibinfo {year} {2005})}\BibitemShut {NoStop}%
\bibitem [{\citenamefont {Falcon}\ \emph {et~al.}(1998)\citenamefont {Falcon},
  \citenamefont {Laroche}, \citenamefont {Fauve},\ and\ \citenamefont
  {Coste}}]{Falcon1998}%
  \BibitemOpen
  \bibfield  {author} {\bibinfo {author} {\bibfnamefont {E.}~\bibnamefont
  {Falcon}}, \bibinfo {author} {\bibfnamefont {C.}~\bibnamefont {Laroche}},
  \bibinfo {author} {\bibfnamefont {S.}~\bibnamefont {Fauve}}, \ and\ \bibinfo
  {author} {\bibfnamefont {C.}~\bibnamefont {Coste}},\ }\href {\doibase
  10.1007/s100510050424} {\bibfield  {journal} {\bibinfo  {journal} {The
  European Physical Journal B}\ }\textbf {\bibinfo {volume} {5}},\ \bibinfo
  {pages} {111} (\bibinfo {year} {1998})}\BibitemShut {NoStop}%
\bibitem [{\citenamefont {Yang}\ \emph {et~al.}(2011)\citenamefont {Yang},
  \citenamefont {Silvestro}, \citenamefont {Khatri}, \citenamefont {{De
  Nardo}},\ and\ \citenamefont {Daraio}}]{Yang2011}%
  \BibitemOpen
  \bibfield  {author} {\bibinfo {author} {\bibfnamefont {J.}~\bibnamefont
  {Yang}}, \bibinfo {author} {\bibfnamefont {C.}~\bibnamefont {Silvestro}},
  \bibinfo {author} {\bibfnamefont {D.}~\bibnamefont {Khatri}}, \bibinfo
  {author} {\bibfnamefont {L.}~\bibnamefont {{De Nardo}}}, \ and\ \bibinfo
  {author} {\bibfnamefont {C.}~\bibnamefont {Daraio}},\ }\href
  {http://www.ncbi.nlm.nih.gov/pubmed/21599325} {\bibfield  {journal} {\bibinfo
   {journal} {Physical Review E}\ }\textbf {\bibinfo {volume} {83}},\ \bibinfo
  {pages} {046606} (\bibinfo {year} {2011})}\BibitemShut {NoStop}%
\bibitem [{\citenamefont {Yang}\ \emph {et~al.}(2012)\citenamefont {Yang},
  \citenamefont {Khatri}, \citenamefont {Anzel},\ and\ \citenamefont
  {Daraio}}]{Yang2012}%
  \BibitemOpen
  \bibfield  {author} {\bibinfo {author} {\bibfnamefont {J.}~\bibnamefont
  {Yang}}, \bibinfo {author} {\bibfnamefont {D.}~\bibnamefont {Khatri}},
  \bibinfo {author} {\bibfnamefont {P.}~\bibnamefont {Anzel}}, \ and\ \bibinfo
  {author} {\bibfnamefont {C.}~\bibnamefont {Daraio}},\ }\href
  {http://linkinghub.elsevier.com/retrieve/pii/S0020768312000534} {\bibfield
  {journal} {\bibinfo  {journal} {International Journal of Solids and
  Structures}\ }\textbf {\bibinfo {volume} {49}},\ \bibinfo {pages} {1463}
  (\bibinfo {year} {2012})}\BibitemShut {NoStop}%
\bibitem [{MC(2013)}]{MC}%
  \BibitemOpen
  \href@noop {} {\emph {\bibinfo {title} {{McMaster-Carr Data Sheet}}}}\
  (\bibinfo  {publisher} {mcmaster.com/standard-borosilicate-glass},\ \bibinfo
  {year} {2013})\BibitemShut {NoStop}%
\bibitem [{Obj(2013)}]{Objet}%
  \BibitemOpen
  \href@noop {} {\emph {\bibinfo {title} {{Objet Printers}}}}\ (\bibinfo
  {publisher} {stratasys.com/materials/polyjet},\ \bibinfo {year}
  {2013})\BibitemShut {NoStop}%
\end{thebibliography}%

\end{document}